\documentstyle[11pt,epsfig]{article}

\def\bc{\begin{center}}
\def\ec{\end{center}}
\def\1#1{{\bf #1}}
\begin{document}

\begin{center}
{\Large \bf Testing the Influence of   Surface Tension and Finite Width of  QGP Bags  on
 the  
 QCD Matter EOS  Properties  at NICA Energies 
}

\vspace{1.0cm}

{\bf Kyrill A. Bugaev
}\\

\vspace{1.cm}

\vspace{0.5cm}

Bogolyubov Institute for Theoretical Physics,\\
Metrologichna str. 14$^B$, 
03680 -- Kiev, Ukraine\\

\hfill \\

e-mail: bugaev@th.physik.uni-frankfurt.de

\date{\today}

\end{center}

\vspace{2.0cm}

\begin{abstract}
Here I  give some strong arguments that the central  issues  for theoretical studies 
of the (tri)critical  endpoint of the QCD phase diagram are the surface tension of large/heavy  QGP bags 
and their medium dependent width. Then I discuss three major directions to  further develop 
the  realistic exactly solvable statistical  models  which simultaneously are able to describe the 1-st order deconfinement phase 
transition, the 2-nd order one and the cross-over.  Also I  analyze  the most necessary projects that have to be 
studied in order to formulate the reliable and convincing signals of the mixed phase formation at NICA
energies.   

\end{abstract}

\vspace{3.cm}


\vspace{2.cm}

\newpage

\section{Motivation}

\vspace*{-0.25cm}

Extensive experimental and theoretical searches for the (tri)critical endpoint of the strongly interacting matter 
became one of the focal points  of the modern nuclear physics. The most powerful computers and very 
sophisticated algorithms are used for the  lattice quantum chromodynamics (LQCD) simulations 
to locate this endpoint and study its properties \cite{fodorkatz,karsch}, but despite these efforts the present situation is far from being clear. 
This fact has a very negative impact on a success of  planned experimental programs 
like the low energy RHIC (BNL), SPS(CERN), NICA (Dubna) and FAIR (GSI), which, in one way or another, are devoted to searches for the QCD phase diagram  endpoint. 
Thus, there is a huge gap in our understanding the physical cause  which is responsible 
for  the QCD (tri)critical endpoint  existence.  As a consequence there is no reliable predictions for 
its location, for  its properties  and for   their   modifications in finite systems which, in fact,  are only available  in the nuclear laboratories.

Searches for the QCD (tri)critical endpoint  are not only the primary goals of the above mentioned experimental programs. In fact,  these programs are aimed at the discovery of the QGP and/or its mixed phase with hadronic matter.  However, at present neither the equation of state (EOS) of strongly interacting matter, nor  an exact location of the deconfinement phase transition (PT) and/or cross-over are known. Despite a limited success,  the present LQCD results at nonzero baryonic densities 
are not very helpful because there are no convincing arguments on the convergency of the existing lattice numerical algorithms and because such algorithms are not  well suited for   the quarks of  physical masses. 
On the other hand  the vast majority  of phenomenological EOS \cite{Misha:07} are not informative because they are based on the mean-field models which are not truly statistical ones and, hence, cannot be reliable both for tracing  the physical mechanism responsible for the QCD (tri)critical endpoint  existence and for the endpoint   properties (its location, critical exponents, PT-order etc). 

Moreover, there exists a huge list of unanswered (and often ignored!)  questions  
related to the following
{\it two main problems} of  modeling  PT  in finite systems which are probed in the nuclear laboratories:  {\bf (1)} how the EOS with PT is  modified in finite systems; and 
{\bf (2)}  how the endpoint properties would look like in a finite system. 
These problems are  routed in the fact that, despite many progress reports
\cite{FiniteSystPT},  there are only a few  general guesses 
how to define PT and the corresponding  analogs of phases in finite systems while the rigorous treatment of this problem is at the initial stage of research   \cite{Bugaev:07rev}. 

In addition, 
these unanswered questions block the necessary  development of
relativistic hydrodynamics which is the main tool  to model PT in heavy ion collisions (HIC). 
The existing hydro models  implicitly assume that the EOS of infinite system may successfully 
describe the phase transformations in a finite system created in HIC. 
The exact analytical  solutions of several statistical models both with a  PT  \cite{Bugaev:07rev,PTfiniteV:1} and without  it \cite{PTfiniteV:2}   found for finite volumes teach us that in this case the analog of mixed phase consist of several metastable states which may transform into each other. Clearly,  the   processes  of  their  transformations  cannot be described by the usual hydro which is dealing with the stable states only. 
In fact, the   main two problems mentioned above prevent  also the development of  the first principle kinetic theory 
of PT in finite systems, which  today   is in a baby-like stage. 

Furthermore,  usually  it is implicitly assumed that the matter created during the HIC is  homogeneous. 
However, the realistic  statistical models of strongly interacting matter  
\cite{QGBSTM,CGreiner:06} tell us    that 
at and above the cross-over  this  matter consists of  QGP bags with the mean volume of several cubic fm. 
Note that  the existence  of QGP bags of  such a volume is supported by the model of QGP droplets  \cite{Droplets}
which successfully  resolved the HBT puzzles at RHIC.
 
Also the assumption that the heavy QGP bags  (resonances) are stable compared to the typical life-time of 
the matter created in  HIC is, perhaps, too strong.  The recent results obtained  within the finite width model 
\cite{FWM}  show us  that  in a vacuum  the mean  width of a resonance of mass $M$   behaves as $\Gamma (M) \approx 600 \left[\frac{M}{M_0}\right]^{\frac{1}{2}}$ MeV (with $M_0 \approx 2.5$ GeV), whereas in a media it grows with the temperature.   
At the moment it is unclear how the finite width of  QGP bags and other implicit assumptions 
affect the accuracy 
of hydrodynamic simulations, but    it is clear that  their a priori  accuracy  cannot be better than 10-15 \%. 
In fact, from the hydro estimates of the HBT radii at RHIC one concludes that, depending on the model, 
the real accuracy could be between 30 \% to 50 \%. 
Clearly,  the same  is true for the hydro-cascade \cite{BD:00,SHUR:01} and  hydro-kinetic \cite{SIN:02} 
approaches. Thus, at present  there are no strong reasons  to believe that  these approaches  are qualitatively 
better than the usual hydrodynamics \cite{HydroFO}. 

Therefore, it turns out  that  investigation of the strongly interacting matter EOS, especially based on exactly solvable models,  is  vitally necessary for a success of  NICA program.  As it is clear from the discussion above 
the realistic EOS  of  strongly interacting matter   obeying the first principles  of statistical mechanics
would  not only  pin down   the (tri)critical  endpoint location, but also should stimulate development
of realistic hydro, kinetic and  first principle hydro-kinetic models \cite{HydroKin} which are  well  suited to  finite systems created in HIC.  

Thus, the exactly solvable models which implement the correct mechanism of the (tri)critical endpoint generation 
are  of principal importance for  theoretical studies of  physics at NICA energies.  I  believe that  the research in this direction will also allow us to formulate the firm signals of the mixed phase 
formation.  It is well known that none of the present  signals, 
the Kink \cite{Kink}, the Strangeness Horn \cite{Horn} and the Step \cite{Step},  
was yet  confirmed by another experimental collaboration. Moreover, these phenomena are explained with 
the hand waving arguments and lack for  a firm theoretical model.

\section{NICA's Special Position,  Main Goals and How to Undertake Them}

It is very hard to  me as a theoretician  to judge the experimental part of  NICA, but I am sure  those
who constructed the Nuclotron back in 90-th, can do almost anything! 
Therefore, I concentrate on the theoretical side of NICA research program. 
It seems that the present situation is very  favorable for a success of NICA project. 
At the beginning of  80-th of the last century there were only guesses of what could the signals 
of the deconfinement PT and no theoretical concept to work them out.  Now,   almost three decades  later 
there are both the vast collection of  the  data measured at AGS (BNL), SPS (CERN) and RHIC (BNL) 
and  the great  amount of theoretical ideas  and  perspective  approaches.  

However,  a few key  theoretical approaches (see the next sections) which are necessary to work out 
the convincing  signals of the deconfinement PT and the mixed phase formation are not completed yet.
Because of  this  the competing low energy programs will hardly be  able to reach their goals prior to  the NICA start. 
This  favorable situation should be used {\it to  establish   the  leadership of NICA project  both in  theoretical and experimental  research  compared to other low energy programs.} 
There are  a few years left
to reach these goals. To do this it is necessary  
to complete  the corresponding  approaches, test them on the existing HIC data and/or on 
the nuclear multifragmentation data and create the necessary theoretical  manpower which will push the  research at NICA.

Working in several leading nuclear laboratories I saw the very same mistakes which, in my mind, strongly 
decelerate their research progress. They are as follows:

\begin{enumerate}

\item
theoreticians from one field do not interact with their colleagues working in the other   field of research
 (which in many respects is similar!);
 
\item
theoretical groups from the same field do not compete with each other, but find the non-interfering research directions;

\item
theoreticians  and experimentalists do not spend sufficient time in discussions and as a result they simply 
do not understand each other. 

\end{enumerate}

JINR, as NICA  host  place,  
has great scientific potential and old academic tradition which hopefully can be used to 
avoid the above mentioned problems and accelerate the research progress. 
Nevertheless, I would like  to stress that for a successful theoretical research  it is  extremely necessary to establish 
the close working contacts 
between  the different theoretical groups or even communities. 

For instance, the group working on  EOS would need to have not only the  close links with the hydro and hydro-kinetic groups, but also with the groups working on different signals because only in this case it will be possible to verify the potential signal on the existing data (prior to the NICA start!). 
Moreover, it is quite  possible that the necessary data were never collected by the  HIC community, but they 
can exist in the nuclear multifragmentation community which for many years studies the nuclear liquid-gas PT and, hence, their  data can be very useful to verify such a 
potential signal.

Below I give the  strong arguments that the {\bf surface tension of the QGP bags} \cite{QGBSTM} and their 
{\bf short life-time}  \cite{FWM,FWM:kin} are 
the key elements that, so far,  being  ignored by other low energy programs, are of crucial importance 
not only for the  correct  EOS formulation and the (tri)critical endpoint location, but also for working  out the firm and convincing signals of
the QGP matter formation.  In fact,  on theoretical side it is necessary to study the whole complex  of 
questions related to these two key issues for an indubitable   success of NICA project.

\section{The Role of Surface Tension in the (Tri)critical Endpoint Existence}

\vspace*{-0.25cm}

Nowadays  the endpoint properties attracted the great attention of HIC community \cite{Misha:07}
which stimulated an appearance of  several  extravagant  models \cite{Goren:05} that  contradict the whole 
concept of critical phenomena. 
Moreover, the general words on similarity with the critical point  properties 
in other substances do not help much because at present there is a single rigorous  theory of critical point in the spin systems  formulated by K. Wilson and collaborators  \cite{Wilson:eps} whereas  the critical point of  real gases and  that one of  nuclear matter 
are, at best rate,  described by the phenomenological statistical models. 
Thus,  the critical point of  real gases is described by 
the Fisher droplet model  (FDM) \cite{Fisher:67, Moretto, Elliott:06}. 
Although the  FDM has been applied to many different systems,
including nuclear multifragmentation  \cite{Moretto}, nucleation of real fluids \cite{Dillmann},
the compressibility factor of real fluids \cite{Kiang}, clusters of the Ising model \cite{Ising:clust}
and percolation  clusters \cite{Percolation}, its phase diagram does not  include  the fluid at all
and,  hence,  is  not theoretically  well defined. 

Furthermore,  although  the statistical multifragmentation model (SMM) \cite{Bondorf:95} is more elaborate and it allows one to 
define the phase diagram of the nuclear liquid-vapor  PT  in the absence of Coulomb interaction of nucleons and without their asymmetry energy  \cite{Bugaev:00,simpleSMM:1}, to  predict the critical (tricritical) endpoint existence for the Fisher exponent $0 < \tau \le 1$ ($1 <  \tau \le  2 $) \cite{Bugaev:00} 
and to  calculate the corresponding critical exponents \cite{Reuter:01},  the location  of  the  SMM (tri)critical  endpoint  at a maximal particle density of nuclear liquid  does not  seem  realistic. 

Moreover,  the relations between the SMM and FDM  critical points are not well established  yet.  Although the Complement method \cite{Complement} allows one to accurately describe the size distribution of  large clusters of 2- and 3- dimensional Ising model within the FDM framework  in a wide range of temperatures, but  a detailed numerical comparison of  the Ising model and FDM  critical endpoints is extremely hard due to large fluctuations even in relatively small systems. On the other hand, in the formal limit, if the eigen volume of the nucleon vanishes, 
the SMM grand canonical partition recovers the FDM partition, but  the analytical properties of  the infinite sums  
generated by the SMM partition derivatives  get changed and, as a result,  the  particle density of  the gaseous phase condensation   is finite not for the Fisher exponent $\tau \le  2$, as in the SMM, but for  $\tau > 2$ \cite{Reuter:01}.
The same reason is responsible for  the  fact that  scaling relations  between the $\tau$ exponent and other critical indices
in  the FDM \cite{Fisher:67}  differ from  the corresponding relations in the SMM   \cite{Reuter:01}.
Thus, despite the formal limit connecting  the FDM and SMM their analytical structure and universality classes are different \cite{Reuter:01}.

The situation with the endpoint of the deconfinement PT is even less clear than that one  in the FDM and SMM. 
The LQCD simulations  of different groups  disagree with each other. 
The  results of   mean-field  models  are   not  informative as well \cite{Misha:07}. 
At the moment  it is even  unclear whether the QCD endpoint is critical or tricritical.
The commonly used Pisarski-Wilczek argument \cite{Rob:84}  based on the universality properties is, strictly speaking, valid for the  chiral restoration PT whereas for the  deconfinement PT it might  be irrelevant.  Furthermore, 
this line of arguments  \cite{Misha:07}  does not explain the  reason  why, depending on the number of flavors, the  finite masses of light quarks demolish  the 1-st or 2-nd order PT  at low baryonic densities and do not destroy  it at high densities.  Therefore, it is necessary to search for the physical  cause which is responsible for a degeneration of the 1-st order deconfinement PT at high baryonic densities  to a weaker PT at the (tri)critical endpoint and to a cross-over at low baryonic densities.

Furthermore, the recent LQCD simulations  \cite{fodorkatz,karsch,LQCD:rev} teach us that  in the cross-over region, even at high temperatures up to $4 \, T_c$  ($T_c$ is the cross-over temperature), the QGP does not consist of 
weakly interacting quarks and gluons and its pressure and energy density are well below of the corresponding quantities of the non-interacting   quarks and gluons. Although such a strongly coupled QGP 
\cite{Shuryak:sQGP} has put a new framework for the QCD phenomenology, the feasibility 
of understanding such a behavior within   the  AdS/CF  \cite{AdS} and/or  within 
statistical models is far from being simple. 

The first steps to resolve these problems are made by formulating the quark gluon bags with surface tension model 
(QGBSTM) \cite{QGBSTM} which enable us  to work out   a unified statistical description of
the 1-st and 2-nd order deconfinement PT  with the cross-over.  This model naturally explains that the reason for 
degenerating  the 1-st  deconfinement PT into a weaker PT at the endpoint and into a cross-over at  low 
baryonic densities is due to negative surface tension coefficient of the QGP bags at high energy densities 
 \cite{QGBSTM}. The  QGBSTM shows that the deconfined QGP phase is just a single infinite bag whereas the cross-over QGP phase consists of  QGP bags of all  possible volumes and only at very high pressures the 
cross-over QGP phase would consist of an infinitely large bag. An important consequence of such a property is that 
the deconfined QGP phase must be separated from the cross-over QGP by another PT which is induced by the change of sign of  the surface tension coefficient of large bags.
Since this additional  PT  exists at zero the surface tension, it is named the surface induced PT \cite{QGBSTM}.   
Furthermore, the QGBSTM teaches us that for the Fisher exponent  the 1-st order deconfinement PT exists for $1 < \tau \le 2 $ only, whereas at the endpoint there  exists  the 2-nd order PT for $ \frac{3}{2} < \tau \le 2 $ and this point is the tricritical one. 

It is necessary to mention that the surface tension of the QGP bags was first estimated long ago
\cite{Jaffe} and its importance for  the PT properties  \cite{Svet:1} is  well known to several communities \cite{Fisher:67, Moretto, Elliott:06, Madsen,Ignat:1}.  However,  the surface tension  significance for 
the QCD (tri)critical endpoint existence  is not 
understood by the HIC community  and  is at the very beginning of exploration. 
Considering  
the  QGBSTM  as a good starting point, it was possible  to formulate  an exactly solvable statistical model for 
the QCD  critical  endpoint, the QGBSTM2  \cite{QGBcrit}. 
These  results  \cite{QGBcrit}  show  that  the  critical  endpoint existence 
requires entirely new statistical approach to generate  the leading singularities of the isobaric partition. 
The model of the critical  endpoint  \cite{QGBcrit}   
also requires the negative value of the surface tension coefficient in the cross-over and for the temperatures above the deconfinement PT.  
Moreover, it was shown that the necessary conditions  for the 1-st order deconfinement PT existence are the discontinuity of the first derivatives of the surface tension coefficient \cite{QGBcrit} and $\tau \ge 2$. Thus,  the QGBSTM2 suggests that the 1-st order deconfinement PT  is the surface induced PT. 
Such a nontrivial structure of the QGBSTM2  explains the reason  why  during almost last three decades  the  formulation of   statistical model  to describe  the QCD  critical  endpoint was unsuccessful.  
Evidently, such a finding along with  the QGBSTM  opens absolutely new possibilities to model 
the  QCD  critical  and  tricritical endpoint.  It seems that the most promising  
\underline{directions  to improve these exactly solvable  models}  are as follows:

\begin{itemize}

\item 
extend and develop the  QGBSTM \cite{QGBSTM} and the QGBSTM2 
\cite{QGBcrit}  in order to normalize them 
onto the existing LQCD data and reproduce the deviation of the QGP EOS  from the Stefan-Boltzmann  limit at zero baryonic densities;

\item 
include the nonzero baryonic and strange chemical potentials  into  the  QGBSTM inspired models  
\cite{QGBSTM, QGBcrit} and test them on the measured  particle yield ratios \cite{PBM}; 

\item 
study  the critical exponents  of the  QGBSTM inspired models  \cite{QGBSTM, QGBcrit}   
and their  possible modifications in finite systems on the basis of the Laplace-Fourier  method 
\cite{Bugaev:07rev,PTfiniteV:1,PTfiniteV:2};

\item
investigate the influence  of the Fisher index $\tau$  on the values of critical indices of 
the  QGBSTM inspired models and  the  possible determination of  $\tau$ value from 
the  cluster scaling picture emerging from  HIC data  \cite{HCluster};

\item
include the Lorentz contraction  of the  excluded volumes \cite{RelVDW:1,RelVDW:2} of light hadrons into the  QGBSTM inspired models  \cite{QGBSTM, QGBcrit}, improve the Van der Waals extrapolations of 
\cite{RelVDW:3} according to  relativistic prescription  \cite{RelVDW:2},  calculate the kinetic coefficients
for  hadronic matter and QGP bags  below and above cross-over and determine the free parameters from the comparison with the AdS/CF results.

\end{itemize}

A comparatively independent direction of  research  is related to the surface tension of the QGP bags. 
The  QGBSTM inspired models  \cite{QGBSTM, QGBcrit}   predict  the dominance  of  non-spherical, or even fractal,  QGP bags above the cross-over transition  which, perhaps,  can  indirectly be  detected  both  in   high energy HIC 
and in the LQCD data. 

Very recently an important  breakthrough was made 
in 
the determination of the QGP bag surface tension coefficient  from the LQCD data  \cite{SurfTension:09}. 
In \cite{SurfTension:09}   we suggest a new phenomenological look at the confinement  which is based on the old MIT  bag model.  Equating the free energies of the confining color string and the elongated cylindrical bag, we got something entirely new: the relation between the string tension, surface tension, thermal pressure and  bag radius. 
The found  relation allows, in principle, 
to determine the bag surface tension directly from the LQCD simulations.
Using this relation we were able to naturally explain the `mysterious', as Edward Shuryak called it 
\cite{Shuryak:08}, maximum in the lattice entropy of the color string.
Also we find out that the {\it QGP bag surface tension is amazingly negative} at the
cross-over region. This is not surprising to us because in our previous works on the (tri)critical endpoint models \cite{QGBSTM, QGBcrit}  we
argued that the only physical reason that the 1-st order deconfinement PT 
 degenerates into the cross-over is just negative value of the QGP bag surface tension  at  the transition temperatures and small values of the baryonic 
chemical potential. Now the LQCD free energy gives us a direct evidence of  the fact that the negative
surface tension is the  only physical reason which transforms the 1-st order deconfinement PT into
a cross-over at low baryonic densities.

These findings suggest that 
it is quite possible that an excess of the surface area of large bags compared to the spherical  ones has  been   already detected  by the  soft photon interferometry  data of the WA98 Collaboration \cite{WA98photons}. 
Up to now such 
 an excess of the soft photons  number \cite{WA98photons} is not well explained  by  rather sophisticated 
models of meson-meson bremsstrahlung \cite{Rapp:softphotons}, but, perhaps, it can be
naturally clarified by an  excess of the surface of  QGP bags compared to the traditional beliefs. 
Indeed, if above the cross-over  each QGP bag has  many  protuberances then, on  the one hand,  their surface should emit 
the black body radiation and, hence, should quickly be  cooled  down  compared to the bag's interior, and, on the other hand, such an 
emission may require an essential  revision of the  HBT results.
Therefore, \underline{the  most favorable  projects  related to the  surface tension of QGP bags} are as follows:

\begin{itemize}

\item 
generalize the Hills and Dales model  \cite{PTfiniteV:2} to the  clusters with the continuous base of deformations and  use it to  estimate  the surface entropy of the QGP bags;

\item 
examine the bimodal \cite{FiniteSystPT,Bugaev:07rev} and 
mono-modal properties of  finite QGP matter with 
positive and negative surface tension coefficient, correspondently, to formulate possible signals 
of the deconfinement  PT, the cross-over transition and  surface induced PT \cite{QGBSTM} in finite systems; 

\item 
study the volume-surface fluctuations of QGP bags \cite{FWM:kin} in the grand canonical, canonical and microcanonical ensembles in order to  reconstruct  an actual  surface  tension coefficient of the QGP bags;

\item 
investigate the HBT radii of highly non-spherical QGP bags above the cross-over and the HBT  emission volume
event-by-event  fluctuations for soft photons and pions  as the possible signals  of  negative surface tension
coefficient;


\item
study the influence  of  shape deformations with positive heights (hills) on the fusion and decay rates of QGP bags to describe the kinetics and hydro-kinetics of the 1-st order PT in finite systems.

\end{itemize}

Investigation of these problems will essentially improve our understanding of the QCD  phase diagram and the 
(tri)critical  endpoint properties in finite systems and, hopefully,  will allow us to find entirely  new signals of the mixed 
phase formation at NICA  energies.  However, such tasks cannot be successfully  solved without accounting for 
the finite width of heavy/large QGP bags, which is discussed in the next section.

\section{The Finite Width of the QGP Bags in Resolving Three Major Conceptual Problems of HIC Phenomenology}

Here I would like to discuss several  issues  that, so far, were ignored by the HIC community and this led to oversimplification of the statistical mechanics of  QGP  bags and, as consequence, to a huge gap in our understanding 
of the HIC process at high energies.

Despite the considerable success of the    statistical  models  discussed above and their remarkable features all of them face {\it three  conceptual problems.}  The first {\it  conceptual problem}  can be formulated by asking a very simple question: 'Why are the QGP bags never directly observed  in the experiments?' The routine argument applied to both high energy heavy ion and hadron collisions is that there exists  a PT  and, hence, the huge energy gap separating the QGP bags from the ordinary (light) hadrons prevents the QGP coexistence at the hadron densities below the PT. The same line of  arguments is also valid  if the strong cross-over  exists. But on the other hand
in the laboratory experiments we are dealing with the finite systems and it is known  from the exact analytical solutions of the  constrained statistical multifragmentation model (CSMM) \cite{PTfiniteV:1}  and gas of bags model (GBM)  
\cite{Bugaev:07rev} that there is a non-negligible probability to find the small and not too heavy QGP bags in thermally equilibrated finite systems  even in the confined (hadronic) phase. 
Therefore,  for finite volume systems  created in  high energy nuclear or  elementary particle  collisions such QGP bags  could appear like any other  metastable states in statistical mechanics,
since in this case  the statistical suppression is just a few orders  of magnitude and not of the order of   the Avogadro number.
Moreover, at the pre-equilibrated stage of high energy collision nothing  can actually prevent their 
appearance. 
The  very same  argumentation is true for the  strangelets 
\cite{Strangelets:A,Strangelets:B} whose intensive searches  
\cite{STRsearches:A, STRsearches:B}
in heavy ion collisions, in  many processes in the  universe and in the cosmic rays 
have not led to any convincing result. 

Then, if  such QGP bags and {strangelets}  can be created there must be a reason which 
prevents their  direct  experimental  detection. 
{As we  showed recently \cite{FWM}  there is an inherent property of the strongly 
interacting matter EOS 
which prevents their appearance 
inside of the  hadronic phase even in finite systems.  
The same  property is also responsible for  the instability 
of  large or heavy  strangelets.

The {\it second conceptual  problem} is
seen  in   a huge deficit of the  number of observed  hadronic resonances \cite{Bron:04}  with masses above 2.5 GeV predicted by the statistical bootstrap model  \cite{SBM} and used, so far,  by all other subsequent  models discussed above. 
Moreover, such a spectrum  has been derived on the basis of such profound models like
the dual resonance model (DRM)  \cite{DRM},  bag model \cite{MITBagM} and GBM 
\cite{Kapusta:81}. Very recently  it  was also shown that the large $N_c$ QCD in 3+1 dimensions  must have a Hagedorn mass spectrum 
\cite{Cohen:09}. 
However,   the modern review of Particle Data Group 
contains  very few  heavier hadronic resonances 
comparing to the expectations of the statistical bootstrap model.
Furthermore, the best  description of particle yields observed in a very wide range of  
collision  energies of heavy ions   is  achieved 
by the statistical model which incorporates  all hadronic resonances not heavier than 2.3 GeV \cite{HG}.
Thus, it looks like that  the  heavier hadronic species, except for the long living ones, are simply absent in 
the experiments \cite{Blaschke:03}. 

In contrast to two previous conceptual problems, the third one is purely theoretical. On the one hand, there exists a
well confirmed high energy hadronic phenomenology based on the Regge poles method \cite{RegBook}.
On the other hand the QCD and QCD inspired approaches are pretty far away from the achievements of this 
particle phenomenology especially concerning the problem of  the  heavy QGP bags width.    
There are many aspects of  a problem how to connect these two fields of research, but from 
practical point of view  it is very necessary to find their  close links  in order to  make the QCD phenomenology 
as powerful as high energy hadronic phenomenology.

Our recent results \cite{FWM,FWM:kin} show that  these  three conceptual problems can be naturally  resolved by the 
inclusion of  the  medium dependent finite width of QGP bags into  the  QGBSTM inspired models  
\cite{QGBSTM, QGBcrit}.  
The finite width model (FWM)  \cite{FWM,FWM:kin}  demonstrates  that the  large width  of the QGP bags  not only explains  the observed deficit in the number of  hadronic resonances, but also clarifies the reason  why 
the heavy QGP bags   cannot be directly observed  as metastable  states in a hadronic phase. 
The FWM  allows one   to estimate the minimal 
value  of the width of QGP bags  being heavier than 2.5  GeV 
 from a variety of the lattice QCD data and get  that 
the minimal  resonance width at zero temperature is about 600 MeV,
whereas  the minimal resonance width at  the  Hagedorn temperature is about 2000 MeV.  
As shown in \cite{FWM}  these estimates are almost  insensitive to the number of the elementary degrees of freedom. 
We analyzed the recent lattice QCD data  and   found that besides $\sigma T^4$   term  the lattice QCD pressure 
contains $T$-linear and $T^4 \ln T$ terms in the range 
of temperatures between 240 MeV and 420 MeV. The presence of the last term in the pressure   practically does not affect the width estimates. 
Also our analysis  shows  that  at  hight temperatures the average mass and width of the QGP bags  behave in accordance with  the upper bound of the Regge trajectory asymptotics (the linear asymptotics) \cite{Trushevsky:77}, whereas at low temperatures they obey  the lower bound 
of   the Regge trajectory asymptotics (the square root one) \cite{Trushevsky:77}. Since the FWM  explicitly contains the Hagedorn mass spectrum,  it allows us to remove  an existing contradiction between the finite number of   hadronic Regge families
and the Hagedorn idea of the exponentially growing  mass spectrum of hadronic bags.

These interesting findings form  a good  starting point  for further exploration.  
Since  the mean width of  large/heavy QGP bags introduces a new and extremely important  scale, it is necessary to 
study its all possible manifestations in  HIC,  since  such a scale defines the in-medium life-time  of QGP bags from
the moment of their formation to the moment of their hadronization \cite{HG,Becattini:1,Bugaev:SPS,Bugaev:RHIC} and freeze-out into gas of free particles. 
Therefore, such a  scale has to be included into any realistic  EOS model,  into any realistic  model for matter  evolution 
\cite{SIN:02,HydroFO,HydroKin}
and into  the realistic  freeze-out model \cite{SIN:02,HydroFO,HydroKin,FO1,FO2}. 
However,  these problems  cannot be solved  immediately since they require a lot of additional work. 
Nevertheless, \underline{the most necessary directions of research related to the finite}
\underline{width of the QGP bags} are as follows:  

\begin{itemize}

\item
estimate the mean width of heavy hadronic resonances from the review of the Particle Data Group;

\item 
consider the large $N_c$ limit of the FWM and establish the  connection between  the FWM  and  the corresponding  limit of  the QCD and QCD  inspired models; 

\item 
elucidate the width of  QGP bags directly  from the LQCD data  of  the metastable branch of the  QGP  EOS;

\item
work out a microscopic  model of sequential decay of heavy QGP bags using approach of  \cite{stringflip,burau} and find the constraints on the width of these bags from the  particle yields data measured in the collisions of elementary particles at high energies
\cite{Becattini:1} and from the HIC data for  the early hadronizing  particles  \cite{Bugaev:SPS,Bugaev:RHIC};

\item 
investigate the processes of melting  and propagation   of QGP bags in the dense and hot media in order to   study  the 
low energy jet  evolution;

\item 
combine all the features of the EOS found while completing the projects related to  the surface tension of the QGP bags.

\end{itemize}


\section{Concluding Remarks}

In the  recent works  \cite{QGBSTM, FWM,  FWM:kin,QGBcrit, SurfTension:09} 
we were able 
 not only   to formulate  the  analytically solvable statistical models for the QCD tricritical 
\cite{QGBSTM} and critical \cite{QGBcrit} endpoint,  
but also  to  push forward an idea  of  the  finite width model of QGP bags 
\cite{FWM,  FWM:kin} and to get a firm  evidence of  negative values of the QGP bag surface tension  at  the transition temperatures and small values of the baryonic 
chemical potential  directly  from the LQCD.
Clearly  such a development 
 improves 
the statistical approach  of  the QGP equation of state 
and brings it forward 
to a 
qualitatively new level of  realism by establishing the Regge trajectories of heavy/large QGP bags both in a vacuum and in a medium 
\cite{FWM, FWM:kin} using the LQCD data.

Such a
coherent  picture 
not only naturally explains the existence 
of  the tricritical  or  critical QCD endpoint  due to the vanishing of  the surface tension coefficient, but 
it
introduces a new time scale into the heavy ion physics and develops  
a novel look at the confinement phenomenology based on the surface tension of 
the QGP bags. 
From these important  findings 
 I conclude   that  their further exploration  will lead to the major shift 
of the low energy paradigm of heavy ion physics and will shed light on the modification of the QCD (tri)critical endpoint 
properties in  finite systems. 

A completion of  this  proposal  will  essentially advance  the  theoretical backup  of  NICA project and 
will allow us to formulate the  firm signals of the mixed phase  formation. Of course, here there are only 
the  most general and/or the most necessary directions  of  research regarding the strongly interacting matter 
EOS and  the important bounds  defined by the spacial and temporal scales  which are put on the QGP bags
by their  surface tension and finite width. 
The more detailed and  specialized suggestions have to be worked out together with the  other EOS groups. 
Further  suggestions on the development of hydro, kinetic and hydro-kinetic  approaches to model the deconfinement PT and cross-over in the actual  HIC require a lot of  additional work and should be formulated in a separate proposal 
after an essential part of  the above formulated tasks are completed.  
The novel  and original  directions of research suggested  here  will definitely  lead to a scientific  leadership and success of the NICA project.





\end{document}